\documentclass[aps,prl,preprint,superscriptaddress]{revtex4}

\usepackage{graphicx}
\usepackage{afterpage}

\begin{document}


\title{Discovery of Griffiths phase in itinerant magnetic semiconductor Fe$_{1-x}$Co$_x$S$_2$}


\author{S. Guo}
\affiliation{Department of Physics and Astronomy, Louisiana State
University, Baton Rouge, Louisiana 70803 USA}

\author{D.P. Young}
\affiliation{Department of Physics and Astronomy, Louisiana State
University, Baton Rouge, Louisiana 70803 USA}

\author{R.T. Macaluso}
\affiliation{Department of Chemistry, Louisiana State
University, Baton Rouge, Louisiana 70803 USA}

\author{D.A. Browne}
\affiliation{Department of Physics and Astronomy, Louisiana State
University, Baton Rouge, Louisiana 70803 USA}

\author{N.L. Henderson}
\affiliation{Department of Physics and Astronomy, Louisiana State
University, Baton Rouge, Louisiana 70803 USA}

\author{J.Y. Chan}
\affiliation{Department of Chemistry, Louisiana State
University, Baton Rouge, Louisiana 70803 USA}

\author{L.L. Henry}
\affiliation{Department of Physics, Southern University, Baton Rouge,
Louisiana, 70813 USA}

\author{J.F. DiTusa}
\affiliation{Department of Physics and Astronomy, Louisiana State
University, Baton Rouge, Louisiana 70803 USA}


\date{\today}

\begin{abstract}
Critical points that can be suppressed to zero temperature are interesting
because quantum fluctuations have been shown to dramatically
alter electron gas properties.
Here, the metal formed by Co doping the
paramagnetic insulator FeS$_2$, Fe$_{1-x}$Co$_x$S$_2$, is demonstrated to 
order ferromagnetically at $x>x_c=0.01\pm0.005$ where we observe unusual transport,
magnetic, and thermodynamic properties. We show that
this magnetic semiconductor
undergoes a percolative magnetic transition
with distinct similarities to the 
Griffiths phase, including singular behavior at $x_c$ and zero temperature.

\end{abstract}


\maketitle


The paramagnetic (PM) to ferromagnetic (FM) transition in magnetic semiconductors is 
a prominent topic in condensed matter physics 
because of efforts to discover materials useful for spintronics\cite{spintrev}. 
Magnetic semiconducting materials are considered essential 
for use as spin injectors
in this nascent technology, yet single phase materials 
with Curie temperatures, $T_C$, well 
above 300 K that are compatible with today's technologies
have not been identified. Problems 
encountered by efforts to increase $T_C$ of Mn doped III-V 
semiconductors suggest that
a deeper understanding is 
needed\cite{matsukura} and recent theoretical investigations 
have provided some progress\cite{priour,schulthess}. 
However, disorder and strong Coulomb interactions
are both central issues in semiconductors with magnetic 
impurities and this makes modeling difficult.
In Mn doped III-V semiconductors, Mn provides localized 
magnetic moments as well as
a smaller number of hole carriers that
couple the local moments via the RKKY interaction.
Models based on a magnetic-polaron Hamiltonian with
random arrangements of RKKY-coupled moments have
predicted a zero-temperature, $T$, percolative transition at critical 
magnetic moment and charge carrier densities\cite{priour}. 
This critical point is governed by the competition between
a nonmagnetic ground state and the magnetically ordered one and because of
the doping induced disorder these materials are expected to display
Griffiths phase singularities.
Other than reports 
in doped LaMnO$_3$ and LaCoO$_4$ suggestive of Griffiths phase 
physics\cite{salamon}, there are no convincing demonstrations of  
Griffiths phases in magnetic semiconductors.
A second class of materials predicted to display the Griffiths phase is the
heavy fermion metals tuned by chemical 
substitution to be near a $T=0$ quantum 
critical point (QCP)\cite{Stewart,Griffiths,CastroNeto,ouyang}.
In this case, rare clusters of strongly coupled magnetic moments 
are predicted to
tunnel between magnetization states over classically 
forbidden regions resulting in non-analytic thermodynamic 
quantities at $T=0$\cite{CastroNeto}.

In this letter we report on
a magnetic and semiconducting system which resembles (GaMn)As
with the advantage that single crystals 
can easily be grown and characterized; Co doped iron pyrite or 'fools gold'.
FeS$_2$ is an insulator with a $\sim 1$ eV band gap.
It is isostructural to CoS$_2$, an
itinerant ferromagnet with $T_C=120$ K\cite{Jarrett}.
FeS$_2$ and
CoS$_2$ form a continuous solid solution over the entire concentration
range, $x$, in Fe$_{1-x}$Co$_{x}$S$_{2}$, where it is
fully
spin-polarized for $0.25< x<0.9$\cite{PyriteHalfMetal}.
Previous magnetic susceptibility, $\chi$, measurements found
FM order at $T > 2$ K for $x\ge 0.05$\cite{Jarrett}. Here, we discover 
metallic behavior for $x\ge 0.001$ with 
magnetic order at $x > x_c = 0.01\pm 0.005$ 
consistent with earlier work. In addition, we observe 
an evolution from a partially Kondo screened metal at $x < x_c$ to a 
ferromagnet characterized by a percolative transition for $x>x_c$.
For $x \approx x_c$ we find a divergent Sommerfeld 
coefficient at low-$T$ indicating non-Fermi liquid behavior along with a
magnetic field $H$ and $T$ dependent magnetization $M$ that suggests
a critical transition to a Griffiths phase-like state. 
In metallic Fe$_{1-x}$Co$_x$S$_2$ we discover evidence for 
the unusual coexistence of magnetically ordered phases and
partial Kondo screening of magnetic moments.
In contrast 
the model\cite{priour} for (GaMn)As predicts a Griffiths phase
under the conditions of localized electronic 
carriers where magnetic-polarons are important. 
In light of the inferred competition between the RKKY coupling and Kondo 
screening, as well as the divergent thermodynamic properties 
at $x_c$ and $T\rightarrow 0$, this system may be more closely described by
models of $f$-electron 
materials displaying quantum criticality\cite{CastroNeto}.

Single crystals were
synthesized by standard I$_2$ vapor 
techniques from high purity starting materials.
Crystals were etched in HCl to remove any remaining flux 
and characterized by single crystal X-ray diffraction
and energy dispersive X-ray microanalysis. The
Co concentration of our crystals is
consistent with the saturated $M$ at 5 T and 
1.8 K and is about 70\% of the nominal concentration of our starting materials. 
$M$ and $\chi$ were measured in a
SQUID magnetometer for $T > 1.8$ K and a
dilution refrigerator  above 50 mK. The resistivity, $\rho$, and 
Hall effect were measured using four-probe lock-in techniques 
at 17 or 19Hz, with thin Pt wires attached using silver paste 
or silver epoxy. The specific heat was measured using a standard thermal relaxation 
method.

Our $M$ measurements identify magnetically
ordered states with $T_C$s
shown in Fig.\ 1a by way of a peak in 
the real part of the AC susceptibility, $\chi'$ (Fig.\ 2a), 
that is
apparent in all of our samples with $x>x_C = 0.01 \pm 0.005$. 
We have checked that $T_c$ signals a FM transition by
comparing to a standard Arrott analysis of
a dense set of $M(H,T)$ data for a few
samples with $x \ge 0.05$\cite{Arrott}. A comparison is made in Fig.\ 1a
between $T_C$ and the Weiss temperature, $\theta_{W}$,
determined from the $T$-dependence of $\chi'$ at $T$'s well above
$T_C$, Fig.\ 2a, b. It is apparent that for $x > x_c$ a FM phase
emerges and that $T_C$ increases systematically with $x$. The scatter
in these data is likely due to the variations in
simultaneously grown crystals as is evident by the range of $\theta_W$
values for crystals of the same or similar $x$. In Fig.\ 1b
the density of  magnetic moments, $n$, determined from fits of the
Curie-Weiss (CW) form (Fig.\ 2b) to 
$\chi(T)$ for $T>>T_C$ are displayed. We have
assumed an effective moment of $J=1/2$ and found 
$n$ much larger than the high-$H$ low-$T$
saturated moment for $x>x_c$. This could indicate 
either a large Rhodes-Wohlfarth ratio as in itinerant 
ferromagnets (between 1.8 and 7 as compared to 3.5 in MnSi), or 
$J > 1/2$ local moments\cite{moriya}.
To investigate this
we have measured the Hall effect
to determine the carrier concentration, $n_{Hall}$, of our
samples. $n_{Hall}$ was determined at
high-$H$ to eliminate anomalous contributions.
Fig.\ 1b shows that $n_{Hall}$ ranges from 10 to 30\% of $x$ indicating
that only a fraction
of dopants donate electrons to a conducting band. Thus, 
it is likely that the difference between the Curie and
saturated moments results from localized electrons with $J > 1/2$.

\begin{figure}[htb]
  \includegraphics[angle=90,width=2.5in,bb=75 350 508 697,clip]{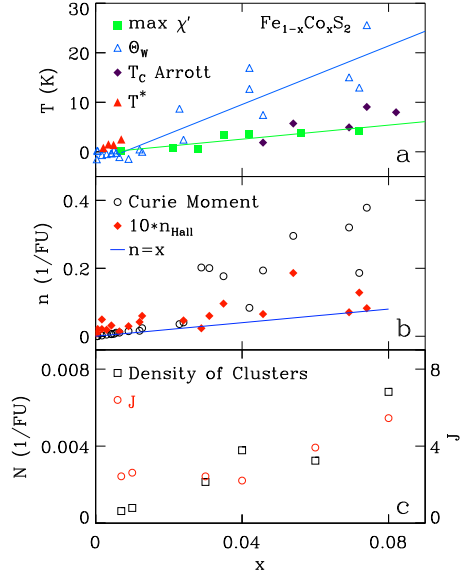}%
  \caption{\label{fig:doping_dependence} 
(Color) Doping dependence (a) $T_C$ from peak in
AC susceptibility, $\chi'$ and from Arrott analysis 
{\protect{\cite{Arrott}}}, Weiss temperature, $\Theta_W$, 
Kondo temperature, $T^*$.
Lines are linear fits.
(b) Curie moment per formula unit from Curie-Weiss analysis 
and Hall carrier density per formula unit, $n_{Hall}$, multiplied by ten.
(c) Density, $N$, and $J$ of spin 
clusters per formula unit.}
\end{figure}

\begin{figure}[htb]
  \includegraphics[angle=90,width=2.75in,bb=60 300 556 696,clip]{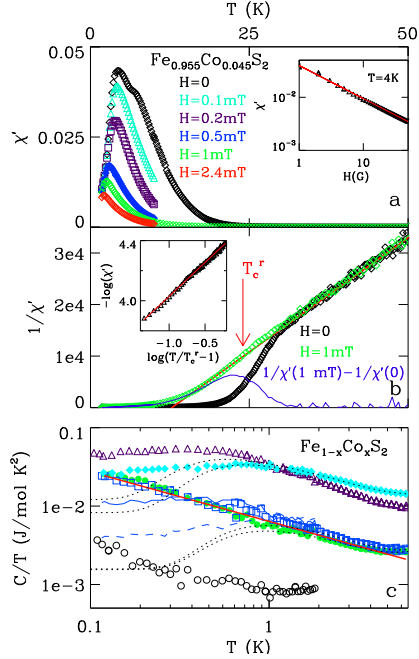}%
  \caption{\label{fig:gamma}
(Color) Susceptibility and Specific heat. 
(a) Temperature $T$ dependence of real part of 
AC susceptibility, $\chi'$.
Inset: $\chi'(H)$ at 4 K with power law fit (red line). 
(b) $T$ dependence of $1/\chi'$ at 0 and 1 mT. Red line 
is fit of Curie-Weiss form at high-$T$.
Inset: $1/\chi'(0)$ vs.\ reduced $T$, ($T/T_C^{r}-1$). 
Red line is $1/\chi'=a(T-T_C^{r})^{\delta}$ with 
$\delta = 0.45\pm 0.04$ and $T_C^{r}=24 \pm 0.5$ K.
(c) Specific 
heat, $C$ divided by $T$, vs.\ $T$ for
$x=0.002$ (circles), $0.005$ (bullets), $0.007$ 
at $H=0$ (blue squares), $H=1$ T (solid line), and $H=3$ T 
(dashed line), $0.03$ (FM) (triangles), 
and $0.045$ (FM) (diamonds).
Red line is fit of the form $aT^{-\alpha}$ 
with $\alpha = 0.69\pm 0.05$, for $x=0.005$.
Dotted lines are fits of a Sommerfeld plus Schottky model
to the data at $T > 2$ K.}
\end{figure}

To explore more fully the properties of 
Fe$_{1-x}$Co$_x$S$_2$, we have
measured the specific heat, $C$, finding nearly identical $C$ 
for all our crystals above 20 K since 
this region 
is dominated by phonons. However, at lower $T$'s we find a 
contribution that grows with $x$ shown in Fig.\ 2c.
Samples that
display a finite-$T$ peak in $\chi$ also display a broad maximum in
$C(T)/T$, albeit at a lower $T$.
The idea of spin clusters suggested by $M(H,T)$ and 
$\chi'(T)$ was probed by comparing $C(T)$ and $\chi'(T)$ to
determine $J$, the fluctuating moment above
$T_C$. At $T>2K$ we fit $C(T)$ by the sum of $\gamma T$, a phonon term
$\beta T^3$ ($\beta=2.17\times 10^{-5}$ J/mol K$^4$),
and a Schottky term due to localized magnetic moments with
$nJ(J+1)$ determined by fits of the CW form to $\chi'(T)$ (Fig.\ 2c).
The best fit $n$ and $J$ are shown in Fig.\ 1c
where we observe that $J>1/2$, $n < x$, and that both grow with $x$.
Thus, spin clusters are consistent with both $C(T)$ above 2 K and
$\chi'(T)$.

The magnetic transitions were explored in detail
by measuring the low-$H$ $\chi'(H,T)$
as in Fig.\ 2a, b. 
What is interesting is that
both the $T$ and magnitude of the $\chi'$ maxima are
significantly suppressed by very small $H$. 
$\chi'(H)$ at 4 K is displayed in the inset to Fig.\ 2a where
a power law form $\chi'(H) = bH^{-\beta}$ with
$\beta=0.62 \pm 0.03$ describes the data well. 
In addition to this extreme field sensitivity we have
observed deviations from CW behavior. In Fig.\ 2b we 
plot $1/\chi'$ at DC fields of 0 and 1 mT
with dramatic changes evident below 30 K. The 1 mT
data follow a CW form, $\chi' \propto (T-\theta_W)^{-1}$
with $\theta_W=13$ K for $T > \theta_W$. 
In contrast, the $H=0$ data cannot be
described by a CW form to much higher $T$s 
and the deviation is in the direction of {\it smaller}
$1 / \chi'$. The growth of $\chi'$ beyond the CW form is 
an indication of short range FM correlations [See e.g.\cite{ouyang}] 
as spin clusters imply larger $\chi'$.
In the inset of Fig. 2b we display a fit of the form
$1/\chi' = (T-T_C^{r})^{\delta}$ to the data, $T_C^{r}=24$
$\pm 0.5$ K 
$\approx 2\Theta_W$ the critical 
$T$ for the largest clusters and $\delta = 0.45 \pm 0.04$,
suggestive of Griffiths phase formation.

Samples that remain PM down to our lowest $T$ 
have a $C(T)/T$ that decreases with $T$ and that can be
accurately described by a $aT^{-\alpha}$ form with $\alpha
<1$. The red line in Fig.\ 2c represents this form 
with $\alpha=0.69 \pm 0.03$. Application of $H>0$
suppresses the low-$T$ $C(T)/T$ of all of our samples and the PM samples
display a $C(T)/T$ that resembles our FM samples.
If we make the assumption
that the conduction electron gas acts
independently from a set of weakly interacting local moments, then 
$C(T)$ would be well fit by the sum of $\gamma T$, $\beta T^3$ and a 
Schottky-like anomaly down to zero-$T$. At finite field, the anomaly would 
evolve into a Schottky peak. Although this description works
well above 1 K, 
our low-$T$ data does not conform to this simple
picture. Interestingly,
materials in proximity to metal-insulator transitions have a diverging
$C(T)/T$ also described by a $T^{-\alpha}$ form.
This is ascribed to
the random position of local moments that are interacting
antiferromagnetically leading to a singlet ground 
state\cite{BhattandLee}. However, since our samples are either
FM or nearly FM, we do not
consider this to be a likely explanation of our data. Instead, we
suggest that the divergence of $C(T)/T$ for $x \sim x_c$
is indicative of Griffiths phase physics and/or to the 
proximity to a FM QCP. 
Magnetic fields or ordering return $C(T)/T$
to a Fermi liquid form at $T$'s proportional to $H$ or $T_C$.

The charge carrier transport properties of 
our crystals 
are presented in Fig.\ 3.\ While
nominally pure FeS$_2$ displays insulating behavior,
Co doping at a level of $x=0.001$ is sufficient to
create a metal (Fig.\ 3a). Larger $x$ tends to increase $n_{Hall}$
and decrease the
resistivity, $\rho$. One
interesting feature is that the $T$-dependence of $\rho$ for 
 $x\approx x_c$ closely follows a $T^{\xi}$ form with $\xi=1.6 \pm 0.1$
over a wide $T$-range similar to that found in MnSi near the
critical pressure for suppression of ferromagnetism. 
It is also consistent with the 
spin fluctuation model of nearly FM metals where a 
$T^{5/3}$ dependence is predicted for itinerant magnets. However,
as noted above, the magnetic moments are likely to result from
localized electrons so 
it is not clear that this model is appropriate.
Evidence for the importance of magnetic 
fluctuations\cite{moriya,millis1,nicklas} in determining
$\rho(T)$ can be seen in Fig.\ 3b where
the $T$-derivative of $\rho$ normalized by $\rho_0=\rho(4 K)$, 
$d\rho/dT /\rho_0$, is displayed.
The normalization removes error in crystal 
geometry determination as well as changes due to variations in $n_{Hall}$.
This quantity is strongly peaked near $x_c$.
Thus, we observe a reduced power-law 
behavior of $\rho(T)$ along with an increased
scattering rate over a wide $T$-range 
in proximity to the zero-$T$ critical point for magnetism. 

\begin{figure}[htb]
  \includegraphics[angle=90,width=3.5in,bb=90 40 532 713]{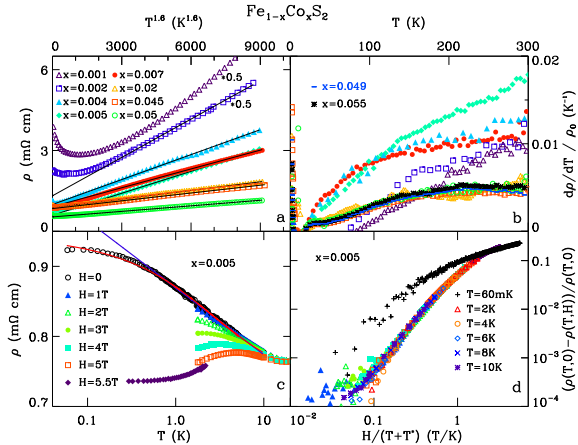}%
  \caption{\label{fig:transport}
(Color) Carrier Transport (a) Resistivity, $\rho$, vs.\ 
temperature to the 1.6 power, $T^{1.6}$, for a subset of our crystals. 
(b) $T$-derivative of $\rho$, $d\rho / dT$, normalized by $\rho(4$ K$)$, 
$\rho_0$, vs $T$.
(c) $T$ dependence of $\rho$ for $x=0.005$ at magnetic 
fields, $H$, identified in the figure.
Blue line is fit of a $\ln{T}$ behavior for $H=0$
and red line is fit of Kondo theory{\protect{\cite{Hamann}}}. 
(d) Scaling plot of magnetoresistance at 
$T$s and $H$s identified in (c) and (d).}
\end{figure}

In addition, a second contribution to $\rho$ is apparent at $T < 20$ K
as demonstrated 
for a PM sample in Fig.\ 3c.
Here $\rho$ increases with
decreasing $T$ in a manner that is well described by a logarithm
over more than a decade in $T$ with Kondo 
temperatures $T^*=$ 0.8, 1.5, 1.4, and 2.5 K for $x=$0.002, 0.004, 
0.005, and 0.007, Fig.\ 1a\cite{Hamann}.
We have also measured a large negative magnetoresistance (MR)
(Fig.\ 3c, d) that is identical in the
transverse and longitudinal current directions.
This indicates a spin, rather than orbital, mechanism
for the MR consistent with a Kondo effect dominating $\rho(T,H)$.
Furthermore, all
of our $T$- and $H$-dependent data can be scaled by
a single ion Kondo form; $\rho(T,H) - \rho(T,0) / \rho(T,0)$ scales as $H /
(T+T^*)$ for $T \ge T^*$ as shown in Fig. 3d\cite{Schlottmann}.
We conclude that a single
energy scale, likely a Kondo coupling of conduction electrons with
the local moments associated with the Co ions, 
determines the low-$T$ $\rho$ of
Fe$_{1-x}$Co$_x$S$_2$.

The observation of power-law divergent $C$ and $\chi'$ for $x \sim
x_c$ with
similar exponents suggests that a single physical mechanism describes
both. The sensitivity to
magnetic fields indicates that there is weak coupling between 
clusters creating a fragile magnetic order for
small $x$. In Fig.\ 4 we summarize the evolution
of Fe$_{1-x}$Co$_x$S$_2$ from a strongly PM low-carrier-density 
metal to a 
FM with increasing $x$. In Fig.\ 4a, for $x<x_c$ and $T=0$, the
system consists of a small density of magnetic moments 
localized on the Co impurity sites. The
electron carriers screen only a portion of these moments
via the Kondo coupling evident in the carrier transport. As $x$
increases to $x_c$ in Fig.\ 4b there is a 
percolative transition at $T=0$ and we observe a divergent $C(T)/T$
at low-$T$. Clearly for a magnetic transition to occur at 
$x \sim 0.01$ 
long range interactions between moments is
necessary. Thus, a percolative transition occurs at $x_c << 0.2$
required for a face-centered-cubic lattice with 
short range interactions\cite{lorenz}. At still larger $x$, 
Figs.\ 4c and d, the FM state
is complete at low-$T$. For $T$s
above the critical point the system is dominated by super-paramagnetic
regions where magnetic moments are correlated, but long range ordering is not apparent.
The system consists of weakly coupled 
clusters that are easily modified by small $H$. In addition, the clusters have a finite
probability to tunnel to nearly degenerate $M$ states.

\begin{figure}[htb]
  \includegraphics[angle=90,width=2.5in, bb=100 75 533 636,clip]{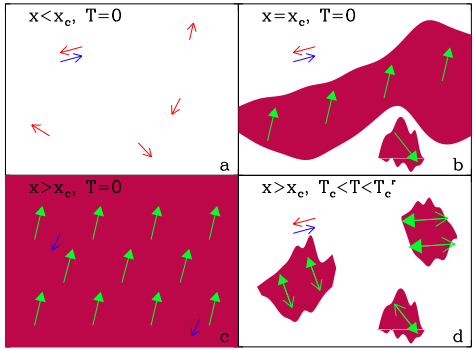}%
  \caption{\label{fig:cartoon1}
(Color) Evolution. (a) Low Co concentration ($x<x_c$)
with disperse local moments (red arrows) and low density of itinerant electrons (blue arrows).
Electrons partially screen moments via Kondo coupling. 
(b) At $x=x_c$ a percolative 
magnetic transition occurs at $T=0$. Ordered regions are
red with magnetization, $M$, direction
green. (c) Larger $x$, $x>x_c$, fully ordered at $T=0$
with large $M$ domains. (d) $T > T_C$, 
clusters of strongly coupled spins form for $T<T_C^r$.
Tunneling of clusters indicated by the double-ended green arrows. 
}
\end{figure}

These features are described by
Griffiths phase models \cite{Griffiths,CastroNeto} where disorder is
sufficient to cause clusters of localized short lived magnetic order at
$T$'s above the global ordering temperature. As the system is
cooled toward $T_C$, these clusters grow and display
switching of $M$ via tunneling. The consequence of this model is a power-law
form, $1 / (T - T_C^{r})^{1-\lambda}$, with $\lambda < 1$, of the
thermodynamic quantities above $T_C$ that can be suppressed with small
to moderate magnetic fields. For Fe$_{1-x}$Co$_x$S$_2$
$0.30 < \lambda < 0.55$. Griffiths phase physics has been suggested
to explain the non-Fermi liquid behavior of heavy fermion
antiferromagnets with N\'{e}el temperatures driven to zero by chemical
substitution. However, Millis {\it et
al.}\cite{millis2} point out that the coupling of magnetic clusters to
conduction electrons will suppress tunneling,
removing the non-Fermi liquid response in the model. In
contrast to well developed conductors,
our materials are nascent metals with poor electrical
screening formed by doping an
insulator. Although Fe$_{1-x}$Co$_x$S$_2$ does not appear to be
described by a magnetic-polaron Hamiltonian as in Ref.\ [3], our 
data suggest that Griffiths phase anomalies, likely influenced by
Kondo screening, are
in fact observable in magnetic semiconductors. The result is
singular behavior at $T=0$ similar to models of Griffiths phases in $f$-electron 
antiferromagnets. 

We thank I. Vekhter and C. Capan for discussions.
JFD, DPY, and JYC acknowledge support of the NSF
under DMR0406140, DMR0449022, and DMR0237664.


\end{document}